\documentclass[twocolumn,prl,showpacs,preprintnumbers,amsmath,amssymb,superscriptaddress,nofootinbib]{revtex4-2}

\usepackage{graphicx}
\usepackage{dcolumn}
\usepackage{bm}
\usepackage{hyperref}
\usepackage[dvipsnames]{xcolor}
\usepackage{tablefootnote}
\usepackage[mathlines]{lineno}

\begin{document}

\preprint{}

\title{Direct mass measurements of neutron-rich zinc and gallium isotopes: an investigation of the formation of the first r-process peak}

\author{Andrew Jacobs}
\email[Current address: Physics Division, Argonne National Laboratory, Lemont,
Illinois, USA; Corresponding author: ]{ajacobs@anl.gov}
\affiliation{TRIUMF, 4004 Wesbrook Mall, Vancouver, BC V6T 2A3, Canada}
\affiliation{Department of Physics and Astronomy, University of British Columbia, 6224 Agricultural Rd, Vancouver, BC V6T 1Z1, Canada}
\author{Stylianos Nikas}
\affiliation{University of Jyvaskyla, P.O. Box 35, Fl-40014 University of Jyvaskyla, Finland}


\author{John Ash}
\affiliation{TRIUMF, 4004 Wesbrook Mall, Vancouver, BC V6T 2A3, Canada}

\author{Behnam Ashrafkhani}
\affiliation{TRIUMF, 4004 Wesbrook Mall, Vancouver, BC V6T 2A3, Canada}
\affiliation{Department of Physics and Astronomy, University of Calgary, 2500 University Drive NW, Calgary, AB T2N 1N4, Canada}

\author{Ivana Belosovic}
\affiliation{TRIUMF, 4004 Wesbrook Mall, Vancouver, BC V6T 2A3, Canada}

\author{Julian Bergmann}
\affiliation{II.~Physikalisches Institut, Justus-Liebig-Universit{\"a}t Gie{\ss}en, 35392 Gie{\ss}en , Germany}

\author{Callum Brown}
\affiliation{Institute for Particle and Nuclear Physics, University of Edinburgh, Peter Guthrie Tait Road, Edinburgh EH9 3FD, United Kingdom}

\author{Jaime Cardona}
\affiliation{TRIUMF, 4004 Wesbrook Mall, Vancouver, BC V6T 2A3, Canada}
\affiliation{Department of Physics and Astronomy, University of Manitoba, 30A Sifton Road, Winnipeg, MB R3T 2N2, Canada}

\author{Eleanor Dunling}
\affiliation{TRIUMF, 4004 Wesbrook Mall, Vancouver, BC V6T 2A3, Canada}

\author{Timo Dickel}
\affiliation{II.~Physikalisches Institut, Justus-Liebig-Universit{\"a}t Gie{\ss}en, 35392 Gie{\ss}en , Germany}
\affiliation{GSI Helmholtzzentrum für Schwerionenforschung GmbH, 64291 Darmstadt, Germany}

\author{Luca Egoriti}
\affiliation{TRIUMF, 4004 Wesbrook Mall, Vancouver, BC V6T 2A3, Canada}
\affiliation{Department of Physics and Astronomy, University of British Columbia, 6224 Agricultural Rd, Vancouver, BC V6T 1Z1, Canada}

\author{Gabriella Gelinas}
\affiliation{TRIUMF, 4004 Wesbrook Mall, Vancouver, BC V6T 2A3, Canada}
\affiliation{Department of Physics and Astronomy, University of Calgary, 2500 University Drive NW, Calgary, AB T2N 1N4, Canada}

\author{Zach Hockenbery}
\affiliation{TRIUMF, 4004 Wesbrook Mall, Vancouver, BC V6T 2A3, Canada}
\affiliation{Department of Physics, McGill University, 3600 Rue University, Montr\'eal, QC H3A 2T8, Canada}

\author{Sakshi Kakkar}
\affiliation{TRIUMF, 4004 Wesbrook Mall, Vancouver, BC V6T 2A3, Canada}
\affiliation{Department of Physics and Astronomy, University of Manitoba, 30A Sifton Road, Winnipeg, MB R3T 2N2, Canada}

\author{Brian Kootte}
\affiliation{TRIUMF, 4004 Wesbrook Mall, Vancouver, BC V6T 2A3, Canada}
\affiliation{Department of Physics and Astronomy, University of Manitoba, 30A Sifton Road, Winnipeg, MB R3T 2N2, Canada}

\author{Ali Mollaebrahimi}
\affiliation{TRIUMF, 4004 Wesbrook Mall, Vancouver, BC V6T 2A3, Canada}
\affiliation{II.~Physikalisches Institut, Justus-Liebig-Universit{\"a}t Gie{\ss}en, 35392 Gie{\ss}en , Germany}

\author{Eleni Marina Lykiardopoulou}
\affiliation{TRIUMF, 4004 Wesbrook Mall, Vancouver, BC V6T 2A3, Canada}
\affiliation{Department of Physics and Astronomy, University of British Columbia, 6224 Agricultural Rd, Vancouver, BC V6T 1Z1, Canada}

\author{Tobias Murb\"{o}ck}
\affiliation{TRIUMF, 4004 Wesbrook Mall, Vancouver, BC V6T 2A3, Canada}
\affiliation{II.~Physikalisches Institut, Justus-Liebig-Universit{\"a}t Gie{\ss}en, 35392 Gie{\ss}en , Germany}

\author{Stefan Paul}
\affiliation{TRIUMF, 4004 Wesbrook Mall, Vancouver, BC V6T 2A3, Canada}

\author{Wolfgang R. Pla{\ss}}
\affiliation{II.~Physikalisches Institut, Justus-Liebig-Universit{\"a}t Gie{\ss}en, 35392 Gie{\ss}en , Germany}
\affiliation{GSI Helmholtzzentrum für Schwerionenforschung GmbH, 64291 Darmstadt, Germany}

\author{William S. Porter}
\affiliation{TRIUMF, 4004 Wesbrook Mall, Vancouver, BC V6T 2A3, Canada}
\affiliation{Department of Physics and Astronomy, University of Notre Dame, Notre Dame, IN 46556, USA}

\author{Moritz Pascal Reiter}
\affiliation{Institute for Particle and Nuclear Physics, University of Edinburgh, Peter Guthrie Tait Road, Edinburgh EH9 3FD, United Kingdom}

\author{Alex Ridley}
\affiliation{TRIUMF, 4004 Wesbrook Mall, Vancouver, BC V6T 2A3, Canada}

\author{Jon Ringuette}
\affiliation{TRIUMF, 4004 Wesbrook Mall, Vancouver, BC V6T 2A3, Canada}
\affiliation{Department of Physics, Colorado School of Mines, Golden, Colorado 80401, USA}

\author{Christoph Scheidenberger}
\affiliation{II.~Physikalisches Institut, Justus-Liebig-Universit{\"a}t Gie{\ss}en, 35392 Gie{\ss}en , Germany}
\affiliation{GSI Helmholtzzentrum für Schwerionenforschung GmbH, 64291 Darmstadt, Germany}
\affiliation{Helmholtz Research Academy Hesse for FAIR (HFHF), GSI Helmholtz Center for Heavy Ion Research, Campus Gie{\ss}en, 35392 Gießen, Germany}

\author{Rane Simpson}
\affiliation{TRIUMF, 4004 Wesbrook Mall, Vancouver, BC V6T 2A3, Canada}
\affiliation{Department of Physics and Astronomy, University of British Columbia, 6224 Agricultural Rd, Vancouver, BC V6T 1Z1, Canada}

\author{Coulter Walls}
\affiliation{TRIUMF, 4004 Wesbrook Mall, Vancouver, BC V6T 2A3, Canada}
\affiliation{Department of Physics and Astronomy, University of Manitoba, 30A Sifton Road, Winnipeg, MB R3T 2N2, Canada}

\author{Yilin Wang}
\affiliation{TRIUMF, 4004 Wesbrook Mall, Vancouver, BC V6T 2A3, Canada}
\affiliation{Department of Physics and Astronomy, University of British Columbia, 6224 Agricultural Rd, Vancouver, BC V6T 1Z1, Canada}

\author{Jens Dilling}
\affiliation{Oak Ridge Natonal Laboratory, 1 Bethel Valley Road, Oak Ridge, TN, 37831, USA}
\affiliation{Department of Physics, Duke University, 120 Science Dr, Durham, NC 27708, USA}

\author{Ania Kwiatkowski}
\affiliation{TRIUMF, 4004 Wesbrook Mall, Vancouver, BC V6T 2A3, Canada}
\affiliation{Department of Physics and Astronomy, University of Victoria, 3800 Finnerty Road, Victoria, BC V8P 5C2, Canada}

\date{\today}

\begin{abstract}
The prediction of isotopic abundances resulting from the rapid neutron capture process (r-process) requires high-precision mass measurements. Using TITAN's on-line time-of-flight spectrometer, first time mass measurements are performed for $^{83}$Zn and $^{86}$Ga. These measurements reduced uncertainties, and are used to calculate isotopic abundances near the first r-process abundance peak using astrophysical conditions present during a binary neutron star (BNS) merger. Good agreement in abundance across a range of trajectories is found when comparing to several metal-poor stars while also strongly deviating from the solar r-process pattern. These findings point to a high degree of sensitivity to the electron fraction of a BNS merger on the final elemental abundance pattern for certain elements near the first r-process peak while others display universality.  We find that small changes in electron fraction can produce distinct abundance patterns that match those of metal-poor stars with different classifications.
\end{abstract}

\maketitle

The formulation of the synthesis of elements heavier than Fe identifies two processes corresponding to neutron captures \cite{B2FH}.  They differentiate between an environment with a low neutron density ($\approx 10^{6} - 10^{10}$ cm$^{-3}$) resulting in a slow capture process (s-process) \cite{RevModPhys.83.157}, and an environment with a high neutron density ($\approx 10^{20}$ cm$^{-3}$) leading to a rapid capture process (r-process) \cite{Horowitz_2019}. Since then, both the s- and r-processes have been separated into `main' and `weak' components with different astrophysical sites correlating to each \cite{prantzos1990s,busso1999nucleosynthesis,winteler2012magnetorotationally,wanajo2013r,wanajo2018nucleosynthesis,siegel2019collapsars}. 

Significant developments have been achieved regarding the r-process following the detection of the Binary Neutron Star (BNS) merger GW170917 \cite{abbott2017gw170817} and its subsequent kilonova AT2017gfo \cite{abbott2017gravitational,kasen2017origin}. From the electromagnetic spectrum, an early blue emission indicates a moderate r-process environment where isotope production up to $A \le 140$ occurred. 
Furthermore, Sr isotopes specifically have been identified in the spectra \cite{watson2019identification} proving the production of light elements around the first abundance peak. The blue emission is followed by a red emission which indicates the production of heavy elements up to lanthanides. Due to the path the r-process takes, nucleosynthesis modeling often requires the use of nuclear models to provide the necessary nuclear inputs where data is lacking. 
As such, it is imperative to provide experimental data for key nuclei along the r-process path, particularly mass values \cite{brettMass2012,AprahamianMass2014,MumpowerMass2015a,MumpowerMass2015b,JiangMass2021}. Conversely, occurring close to the `valley of beta stability' where almost all needed nuclear data is available, the s-process is comparatively well understood \cite{Reifarth_2014}. 

For elements in the second and third r-process peaks, an abundance universality has been found \cite{cowan2021origin}. This universality is indicated by a small spread in the relative abundance pattern for elements ranging $56 \leq Z \leq 79$ and furthermore matches the solar system's r-process abundance pattern. If the same relative abundance pattern is applied to elements below the second r-process peak, there is a significantly larger scatter indicating a lack of universality. Recently, evidence suggest that a degree of universality exists for some elements between the first and second r-process peaks when normalized to Zr instead of Eu \cite{roederer2022uni}. This points to a potential decoupling of the main and weak r-process. Of the elements that do no exhibit a universality in the region, those from $44 \leq Z \leq 50$ can be explained due to fission of elements heavier than U in the main r-process \cite{roederer2023element}. Anomalies in the Ge, As, and Se abundances have been detected in certain stars which point to an additional intermediate (or i-process) neutron capture \cite{roederer2016diverse}. Additional studies of Carbon Enhance Metal Poor stars has produced conflicting results regarding the need for an exotic i-process \cite{Karinkuzhi2021Low-mass,Hansen2023Evidence,Mashonkina2023PIGS}. As such, further investigation of this region is necessary.

In simulations of potential r-process astrophysical sites, it has been found that light and heavy r-process elements can be produced in distinct ejecta from the same site \cite{wanajo2014production,just2015comprehensive,nishimura2015r,radice2016dynamical,fujibayashi2023comprehensive}. Therefore, this decoupling points to the possibility of investigating the nature of the weak r-process in a BNS merger for elements $32 \leq Z \leq 42$.

In this Letter we report the first direct mass measurements of $^{83}$Zn and $^{86}$Ga as well as an improved mass uncertainty of $^{85}$Ga. We revisit the formation of elements from $32 \leq Z \leq 42$ with state-of-the-art r-process nucleosynthesis calculations in an astrophysical environment conducive to that of a BNS merger. We show that the blue kilonova can reproduce a robust weak r-process pattern for elements which exhibit universality while simultaneously replicating the anomalous spread in Ge, As, Se, and Sr.

High precision mass measurements were performed at TRIUMF's Ion Trap for Atomic and Nuclear science (TITAN) \cite{Dilling2003TheIsotopes} located in the Isotope Separator and ACcelerator (ISAC) \cite{Ball2011PhysicsTRIUMF-ISAC} facility at TRIUMF in Vancouver, Canada. This facility provides access to short-lived radioactive isotopes involved in these processes. In this Letter, we report the first science results from the novel target developed at TRIUMF-ISAC utilizing a W-neutron converter \cite{Egoriti_2018}.Resonant laser ionization \cite{Lassen2017} was used for Zn, and standard surface ionization was employed for other species \cite{Dombsky2000}. The radioactive ion beam (RIB) was reduced to a single A/q by a magnetic dipole mass separator \cite{bricault2002triumf} and delivered to TITAN to perform high precision mass measurements using the MR-TOF-MS.

The TITAN Multiple-Reflection Time-of-Flight Mass Spectrometer (MR-TOF-MS) \cite{Jesch2015TheTRIUMF,REITER2021165823} is based on the Gie{\ss}en-GSI design used at the FRS Ion-Catcher \cite{PLA2013134,DICKEL2015172,DICKEL2017TFS}.  
The measurements were performed following an established procedure \cite{leistenschneider2018dawning,REITER2021165823}. As a preparation step, the potentials of the MR-TOF-MS mirror closer to the injection trap can be lowered, and the ions are dynamically re-captured in the injection trap. This process, called mass selective re-trapping \cite{dickel2017isobar,Jacobs2023Thesis,Jacobs2023ReTrap}, enables the suppression of isobaric contamination by up to four orders of magnitude and has been utilized in several previous experiments at TITAN \cite{Beck2021,Mukul2021,Izzo2021,Porter2021}.

\begin{figure}[tb]
    \centering
    \includegraphics[width=0.49\textwidth]{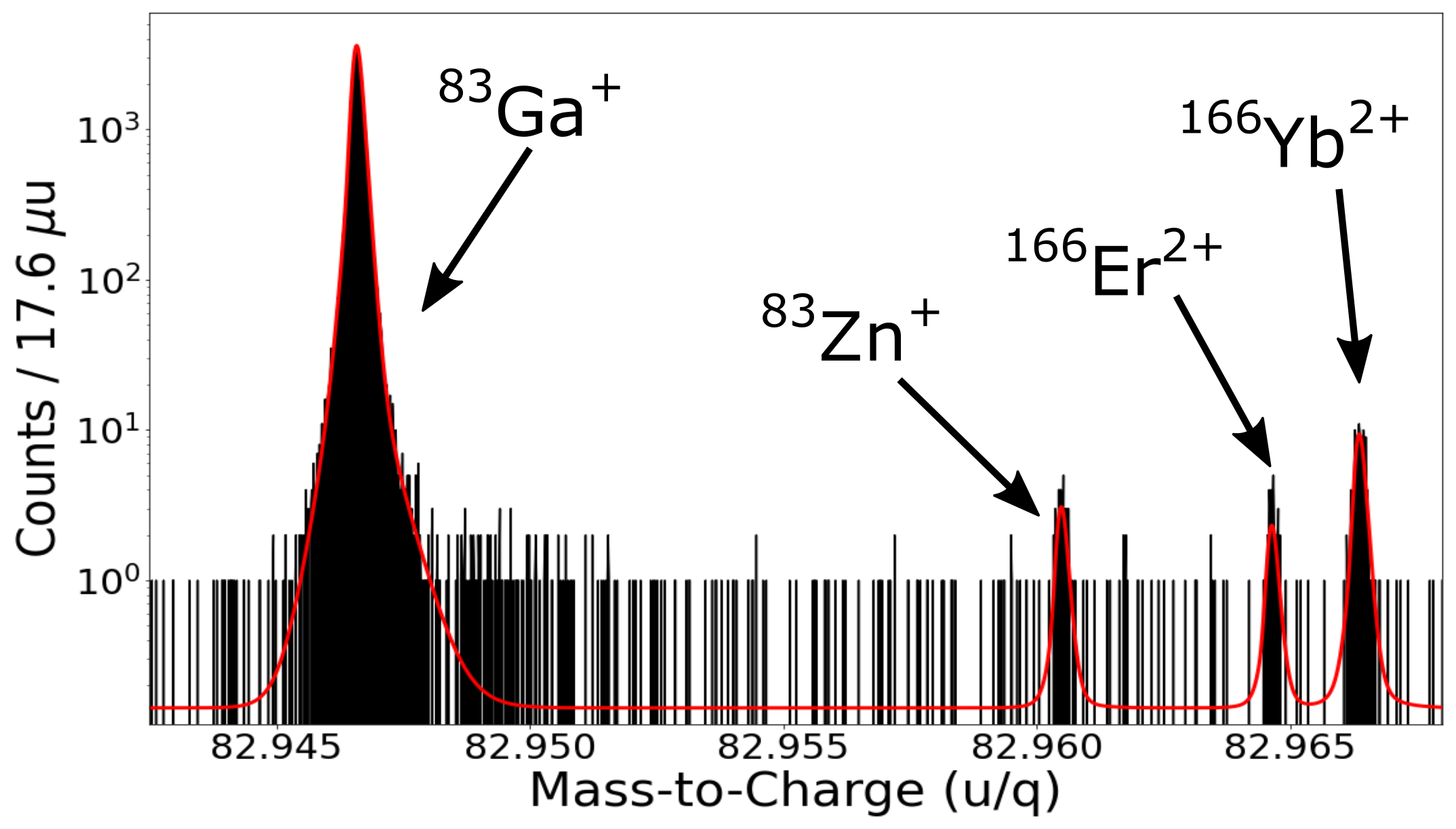}
    \caption{The spectrum and fit (red) for A/q = 83 at 900 turns after 2.5 hours accumulation. $^{83}$Ga$^{+}$ has been used as the calibration species.}
    \label{fig:Spectrum}
\end{figure}


\begin{table*}
\centering
\begin{tabular}{c c c c c c}

\hline \hline \\ [-1em]
Species & Calibrant & Mass Ratio & ME$_{\textrm{TITAN}}$ (keV/c$^{2}$) & ME$_{\textrm{AME2020}}$ (keV/c$^{2}$) & Difference (keV/c$^{2}$)\\
[-1em]\\
\hline \\ [-1em]
$^{79}$Zn & $^{79}$Ga  & 1.000136781 (51) & -53430.0 (79) & -53432.3 (22) & 2.3 (8.4)\\
[-1em]\\
$^{79m}$Zn & $^{79}$Ga & 1.00013678 (16) & -52491 (14) & -52330 (150)$^{*}$ & 161 (151) \\
[-1em]\\
$^{80}$Zn & $^{80}$Rb & 1.000275561 (48) & -51661 (10) & -51648.6 (26) & -12 (11)\\
[-1em]\\
$^{81}$Zn & $^{81}$Ga  & 1.00015146 (11) & -46209 (12) & -46200 (5) & -9 (13)\\
[-1em]\\
$^{82}$Zn & $^{82}$Ga  & 1.000139116 (91) & -42312 (10) & -42314 (3) & 2 (10)\\
[-1em]\\
$^{83}$Zn & $^{83}$Ga  & 1.00016756 (22) & -36311 (18) & \#\#\# & \#\#\# \\
[-1em]\\
$^{85}$Ga & $^{85}$Rb  & 1.00053666 (15) & -39720 (14) & -39740 (40) & 20 (42)\\
[-1em]\\
$^{86}$Ga & $^{86}$Rb  & 1.00061203 (23) & -33768 (20) & \#\#\# & \#\#\# \\
[-1em]\\
\hline \hline
\end{tabular}
\caption{Ga and Zn mass measurements with their respective calibrants, mass ratios ($m_{species}/m_{cal}$), and mass excess results (ME$_{\textrm{TITAN}}$) compared to the most recent AME2020 mass excess (ME$_{\textrm{AME2020}}$). All species were measured in the singly charged (1+) state. Mass Excesses (ME) and mass ratios are reported as atomic masses. Previously unmeasured species are marked as \#\#\# . $^{*}$The recently published \cite{nies2023further} mass excess is reported as -52490 (12) keV/c$^{2}$ resulting in a difference of -1 (18) keV/c$^{2}$.}
\label{tab:Mass}
\end{table*}

To extract the masses of the species of interest, a data analysis procedure using the methodologies described in \cite{AyetSanAndres2019,Paul2021Ga} was performed. A resulting spectrum in which the mass of $^{83}$Zn was determined can be seen in figure \ref{fig:Spectrum}. To maximize the resolving power of the system, a time-resolved calibration (TRC) is done \cite{AyetSanAndres2019}. For the first time, a newly developed ion manipulation technique referred to as `post re-trapping beam merging' was used to ensure that a TRC species was available, regardless of RIB composition. This technique is performed between the isobar separation and mass measurement steps. After re-trapping, the isobarically purified ion ensemble is mixed with a controlled amount of calibration ions from an off-line ion source. The now merged groups are then injected into the TOF analyzer for a mass measurement cycle.
 
In this particular experiment, post re-trapping beam merging was used to achieve a sufficient resolving power needed to separate $^{86}$Rb$^{+}$ and $^{86m}$Rb$^{+}$ via a $^{85}$Rb$^{+}$ TRC. As such, an isobaric calibration species is provided for $^{86}$Ga$^{+}$. This, in turn, made the first high-precision measurement of $^{86}$Ga$^{+}$ possible. Details on this novel technique are described in \cite{Jacobs2023Thesis,Jacobs2023ReTrap}.

The \textsc{emgfit} Python package \cite{paul_stefan_f_2020_4731019} was then used to fit the peak centroids using a hyper-exponentially modified Gaussian lineshape \cite{PURUSHOTHAMAN2017245}. From the centroids, a calibration species with a previously precisely measured mass is used to extract the final mass of the species of interest. Statistical and systematic uncertainties are determined using the procedures described in \cite{AyetSanAndres2019,Paul2021Ga}. Ultimately, a systematic uncertainty of $\delta m / m = 1 \cdot 10^{-7}$ has been determined for these measurements resulting in absolute uncertainties of $\delta m < 20$ keV/$c^{2}$.

Table \ref{tab:Mass} provides a list of the measured masses for $^{79-83}$Zn and $^{85,86}$Ga. The calibration species as well as the mass ratios are reported to account for any adjustment in the calibrant masses in the future. The mass excess results (ME$_{\textrm{TITAN}}$), previous high precision measurements (ME$_{\textrm{AME2020}}$) \cite{Wang2021TheReferences}, and the difference between the two are reported. 

All results are in good agreement with literature values. The uncertainty of $^{85}$Ga, reported by TITAN in \cite{reiter2020mass}, has been reduced by a factor $\approx$ 3 and is in agreement with the previous value. The excitation energy of $^{79m}$Zn is in agreement with the recent ISOLTRAP and JYFLTRAP measurements \cite{nies2023further}. Lastly, first direct measurements of $^{83}$Zn and $^{86}$Ga have been performed. 

To interpret the impact of these results, we use the statistical model code TALYS-1.8 \cite{Kon08a,KONING20122841} to calculate the corresponding (n,$\gamma$) reaction rates with the new mass values. Two different mass tables were used. The first mass table used experimental nuclear masses from AME2020 and, where experimental data is unavailable, the FRDM2012 mass tables \cite{FRDM2012} (referred to henceforth as AME+FRDM). The second mass table used the same combination of AME2020 and FRDM2012, but included the new and improved TITAN mass values in place of either AME2020 or FRDM2012 values (referred to as AME+FRDM+TITAN).

The transmission coefficients for the formation of the compound nucleus are calculated from a spherical neutron-nucleus optical potential using the Koning-Delaroche globally fitted parameters \cite{Kon03a}. The Back-Shifted Fermi Gas model is used \cite{Dil73a,Gil65a} to describe level densities, and the Kopecky-Uhl generalized Lorentzian \cite{Kopesky-Uhl} is used for the $\gamma$ strength functions. The reaction rates were calculated for each mass table using the corresponding central mass values to calculate the reaction Q-values. In order to determine the mass uncertainty's contribution to the final abundance uncertainty, a set of 100 random variations of the masses within the $1\sigma$ uncertainty band of known masses and the reported error of $\sigma_{th} = 0.5595$ MeV from FRDM2012 \cite{FRDM2012} were generated in order to probe the effect of the isotope's mass uncertainty for each astrophysical trajectory. 

The nuclear reaction network code GSINet was used to model the r-process and calculate the elemental  abundances \cite{Mendoza2015GSINet}. We initialized calculations at Nuclear Statistical Equilibrium (NSE), setting the initial temperature as $T_0 = 10$ GK. Due to the uncertainties in the astrophysical conditions that could produce nuclei in the first r-process peak region ($A\approx82$), we simulated nucleosynthesis for a set of parameterized r-process trajectories where we treat initial electron fraction ($Y_e = N_{p} / (N_{p} + N_{n}$)) as a free parameter to investigate. This range is encompased by expected electron fractions found in BNS mergers.

The expansion time scale was set as $\tau = 7$ ms, and the initial specific entropy as $15\ k_B/$ baryon. We assume that the density evolution of the ejecta initially follows an exponential expansion and at later times a homologous expansion \cite{lippuner2015r}.We evolve our abundances assuming NSE for $T \geq 6$ GK, before changing to the full nuclear network calculations.  

Previous work near the first r-process peak has shown a good agreement between parts of the solar r-process abundances and predictions from network calculations \cite{reiter2020mass}. However, due to its age, the Sun has likely undergone several enrichment events \cite{goriely1999uncertainties}. Additionally, the underlying assumptions which subtract only the s- and p-processes could be insufficient \cite{bisterzo2017galactic,cote2018process}. As such, metal-poor (i.e. old) stars offer additional insights due to experiencing fewer enrichment events. In this work we explore the abundances of r-process enriched stars from \cite{roederer2022uni} as well as the r/s-star HD94028 \cite{roederer2016detailed} which has been a motivating factor in probing the i-process. 

\begin{figure}
    \centering
    \includegraphics[width=0.48\textwidth]{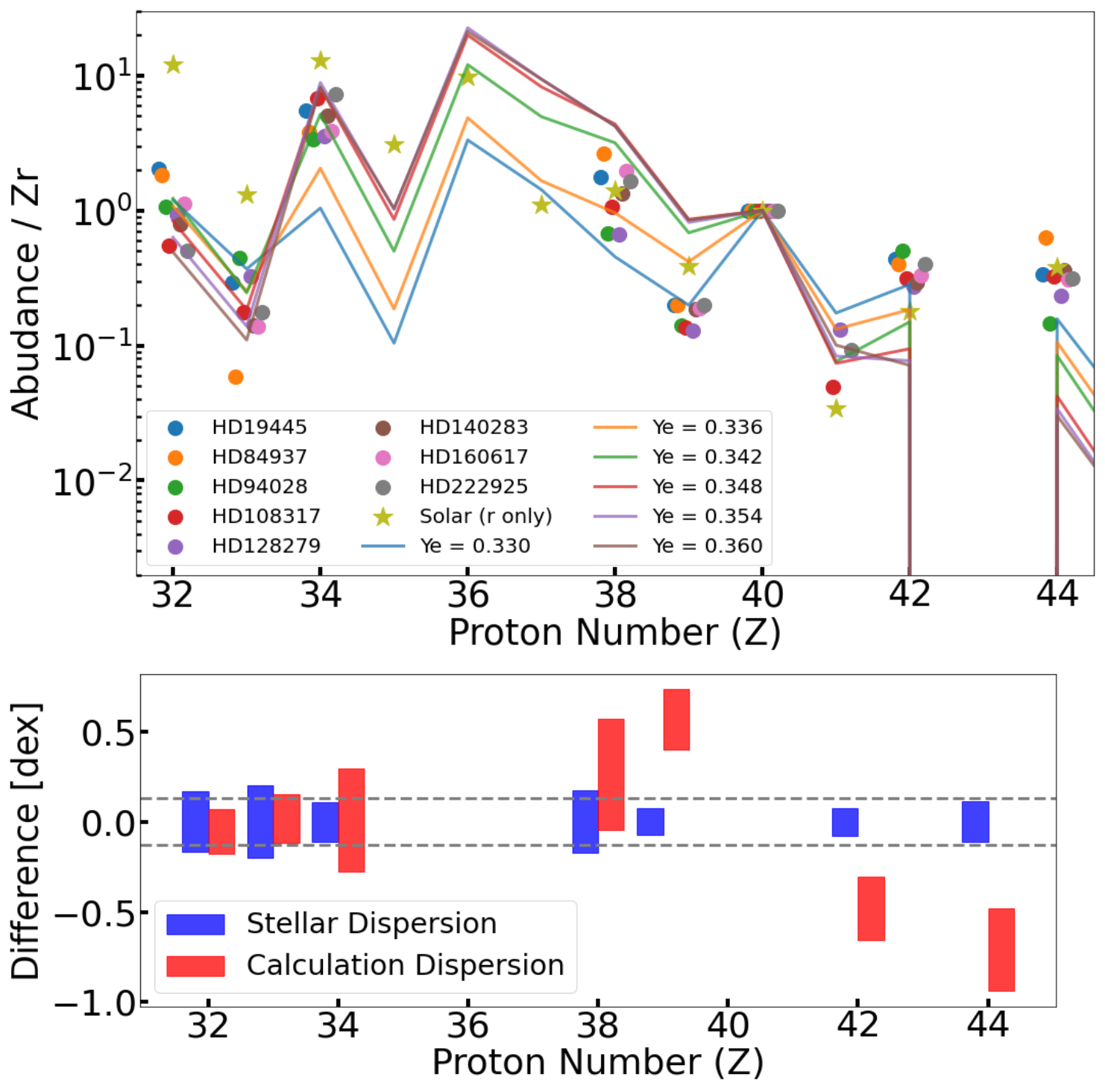}
    \caption{(Top) Calculated elemental abundances of the 1st r-process peak for a wide range of $Y_{e}$ normalized to Zr using AME2020 and new TITAN mass data compared to the abundance patterns of the stars used in \cite{roederer2022uni}, HD94028 \cite{roederer2016detailed}, and the solar r-process pattern \cite{prantzos2020chemical}. (Bottom) The difference (in dex) between the dispersion of the stellar population and network calculations using the new TITAN results for the range of $Y_{e}$ above.}
    \label{fig:ZrAbu}
\end{figure}

Following the calculation of each elemental abundance, the results were normalized for each value of $Y_{e}$ to the abundance of Zr ($Z=40$). To allow comparison, this procedure was also applied to the measured abundance patterns of all stellar data. In Figure ~\ref{fig:ZrAbu}, the resulting abundances for a selection of $Y_{e}$ is shown from $Z=32-44$ compared to the observed abundance patterns. It can be seen that the solar r-process pattern strongly deviates from the distribution of metal-poor stars for several elements.
By varying $Y_{e}$ in a narrow range (which is within the range of $Y_{e}$ estimated from a blue kilonova \cite{wanajo2018physical,wu2019fingerprints}), the resulting TITAN abundance pattern can match the varying abundance patterns of the metal-poor stars. However, a strong odd-even effect from $Z=38-40$ is not reproduced for any value of $Y_{e}$, and abundances for Mo and Ru ($Z=42$ and 44 respectively) are slightly under-predicted.

In the lower plot of figure \ref{fig:ZrAbu}, the dispersion of the stellar data and TITAN calculations is shown. The difference in the mean values between the stellar and TITAN results are calculated as $\textrm{log}(\overline{X}_{\textrm{stellar}} / \overline{Zr}_{\textrm{stellar}}) - \textrm{log}(\overline{X}_{\textrm{TITAN}} / \overline{Zr}_{\textrm{TITAN}})$, while the dispersion is calculated as the standard deviation. The dispersions and/or mean values match well for many elements with the $Z=38-40$ region again providing a source of tension.

\begin{figure}
    \centering
    \includegraphics[width=0.48\textwidth]{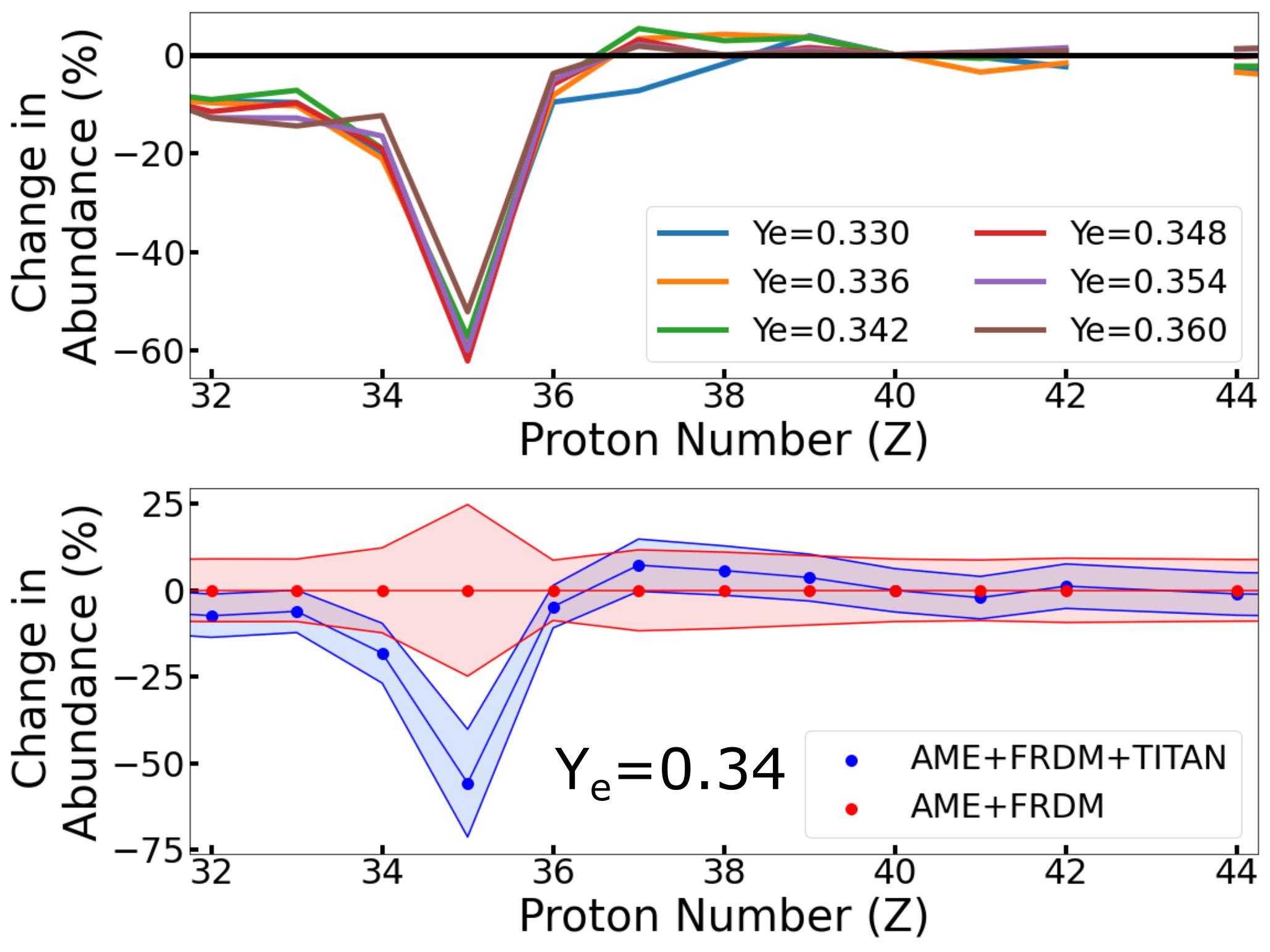}
    \caption{(Top) The difference in abundance between the AME+FRDM and AME+FRDM+TITAN datasets for a range of $Y_{e}$ matching that seen in figure \ref{fig:ZrAbu}. (Bottom) The percent difference between AME+FRDM and AME+FRDM+TITAN results for $Y_{e} = 0.34$ and their corresponding relative uncertainties to 1$\sigma$.}
    \label{fig:abu_cal_comp}
\end{figure}

In Figure \ref{fig:abu_cal_comp}, the abundance results are compared between the two different datasets (AME+FRDM and TITAN). In the upper panel, the differences in abundance between AME+FRDM and TITAN are shown for the same $Y_{e}$ values as those seen in Figure \ref{fig:ZrAbu} (again normalized to Zr). The updated TITAN mass values predict a decrease in $32 \leq Z \leq 35$ with Br having the most significant drop. Other elements remain relatively unaffected. While no Br abundances have been measured outside of the solar system, this reduction in Ge, As, and Se bring the calculations into better general agreement with the metal-poor stars. The agreement for Ge is slightly worsened by $\approx$ 5\%, and the agreement for As and Se is improved by 11\% and 21\% respectively. Furthermore, the spread in distribution due to varying $Y_{e}$ is reduced by $\approx$ 10\%. In the lower panel, the change in abundance is shown for $Y_{e} = 0.34$ with the respective uncertainties of both datasets calculated from the Monte Carlo methods described above. The new mass values reduce the abundance uncertainty due to the nuclear masses by a factor of 2 for $Z<40$.

\begin{figure}
    \centering
    \includegraphics[width=0.48\textwidth]{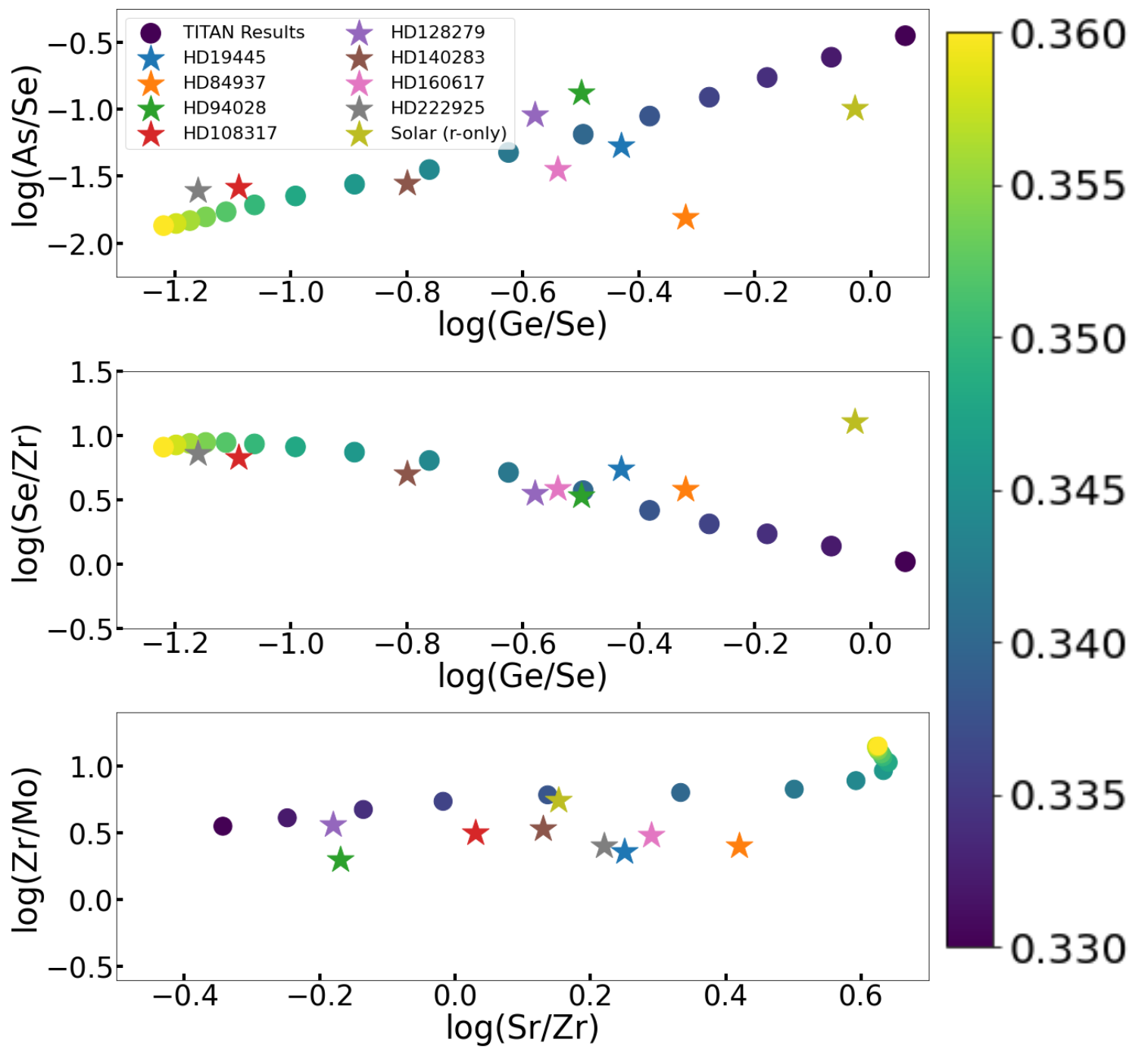}
    \caption{Comparison of logarithmic ratios of key element pairs between TITAN results (with an additional color scale axis for $Y_{e} = 0.33-0.36$ in steps of 0.002) and observations of various metal-poor stars. The trends in ratios from observational stellar data matches that of the $Y_{e}$ range produced by a blue kilonova.}
    \label{fig:Abu_Ratios}
\end{figure}

Based on both observational data and simulation work a spread in both $Y_{e}$ in r-process events as well as in final elemental abundance distributions is to be expected. Therefore, model predictions should closely match the trends between element pairs near the first r-process peak. In figure \ref{fig:Abu_Ratios}, our calculations are compared with those of the metal-poor stars in figure \ref{fig:ZrAbu}.

For the pair Ge/Se and As/Se, a positive linear relationship is expected spanning approximately an order of magnitude for each ratio. The TITAN results match this trend exceptionally well. For the pair Ge/Se and Se/Zr, a negative approximately linear relationship is expected. Again, our calculations align with expectations. Lastly, for the pair Sr/Zr and Zr/Mo, a flat relationship is expected (i.e., as the ratio Sr/Zr changes there should be no change in Zr/Mo). While our absolute value of Zr/Mo is slightly off, our calculations predict a flat relationship between the ratios. Thus, not only do our calculations project the correct spreads in abundance for non-universal elements near the first r-process peak, they also are capable of recreating key elements that reflect universality in this region. As such, we conclude that our weak r-process calculations are capable of explaining the general distribution of elemental abundances from $32 \leq Z \leq 42$ for a variety of stars.

In summary, we have performed high precision mass measurements of neutron rich Zn and Ga isotopes including first time measurements of $^{83}$Zn and $^{86}$Ga. These measurements were enabled by the development of the new ISAC proton-to-neutron converter target as well as the MR-TOF-MS beam merging technique. With these new mass values, state-of-the-art r-process network calculations were performed for trajectories conducive to the astrophysical environment of a BNS merger's neutrino drive winds which produce a weak r-process. We find that, by varying the initial electron fraction, $Y_{e}$, we can reproduce the abundance patterns of a variety of metal-poor stars. Furthermore, we find that by varying $Y_{e}$, we reproduce the correlations between key elemental ratios of these metal-poor stars which show both universality and variations. Ultimately, our results point to small differences in $Y_{e}$ at a given astrophysical site producing large observed deviations for certain elements and little to no change for other elements. A consequence of this result is the calling to question the explicit need for an i-process to explain the observed abundances in a star such as HD94028. To further test these conclusions, additional abundance observations of the first r-process peak in metal-poor stars (particularly such as those in Reticulum II \cite{ji2016r}) would prove useful.

We would like to thank the Targets and Ion Source group at TRIUMF. Particularly, thank J. Lassen for the Zn laser development as well as A. Gottberg for the proton-to-neutron converter target development. Additionally, we would like to thank N. Vassh for fruitful discussion pertaining to our results. This work was supported by the Natural Sciences and Engineering Research Council (NSERC) of Canada under Grants No. SAPIN-2018-00027, No. RGPAS-2018-522453, and No. SAPPJ-2018-00028, the National Research Council (NRC) of Canada through TRIUMF, the Canada-UK Foundation, the UK Science and Technology Facilities Council (STFC) No. ST/V001051/1, German institutions DFG (grants FR 601/3-1, contract no.\ 422761894 and SFB 1245 and through PRISMA Cluster of Excellence), BMBF (grants 05P21RGFN1 and 05P24RG4), and by the JLU and GSI under the JLU-GSI strategic Helmholtz partnership agreement. 



\bibliography{ref}

\end{document}